\documentclass[aps,prl,twocolumn,10pt,superscriptaddress,nofootinbib,balancelastpage]{revtex4-1} 
\usepackage[latin1]{inputenc}
\usepackage{amsmath,amssymb}
\usepackage{mathrsfs} 
\usepackage[capitalise]{cleveref}
\usepackage{siunitx}
\usepackage{braket}
\usepackage{calc}
\usepackage{tabularx}
\usepackage{dsfont}
\usepackage{ifthen}

\allowdisplaybreaks

\newcommand{\ii}{\mathrm{i}}%
\newcommand{\dif}{\mathrm{d}}%
\newcommand{\norm}[1]{\lVert#1\rVert}%
\newcommand{\Tr}{\operatorname{Tr}}%
\newcommand{\ZT}[1]{\textquotedblleft#1\textquotedblright}%

\newcolumntype{Y}{>{\centering\arraybackslash}X}%
\newcolumntype{Z}{>{\raggedright\arraybackslash}X}%

\newlength{\myl}%
\newcommand{\SUM}[2]{{\setlength{\myl}{\widthof{$\displaystyle\sum_{#1}^{#2}$}*\real{0.5}-\widthof{$\displaystyle\sum$}*\real{0.5}}\sum_{#1}^{#2}\;\hspace{-\the\myl}}}
\newcommand{\INT}[3]{\settowidth{\myl}{$\displaystyle\int_{#1}^{#2}$}{\int_{#1}^{#2}\;\;\;\hspace{-\the\myl}\dif #3}\,}
\newcommand{\TINT}[3]{\settowidth{\myl}{$\int_{#1}^{#2}$}{\int_{#1}^{#2}\!\ifthenelse{\equal{#1#2}{}}{}{\;\;\;\;\hspace{-\the\myl}}\dif #3}\,}%
\newcommand{\EINT}[3]{\settowidth{\myl}{$\int_{#1}^{#2}$}{\int_{#1}^{#2}\;\;\;\,\hspace{-\the\myl}\dif #3}\,}

\newcommand{\msc}{\!\neq\uu}%
\newcommand{\mm}{\Gamma_{\msc}}%
\newcommand{\vmm}{\vec{\Gamma}_{\msc}}%
\newcommand{\uu}{\vec{u}}%
\newcommand{\INTOII}{\INT{S_{2}}{}{\Omega}}%
\newcommand{\ff}[3]{f_{#3}^{(#1\mathrm{D})}}%
\newcommand{\TT}[3]{\mathrm{T}_{#3}^{(#1\mathrm{D})}}%

\begin{document}
	
\title{Orientational order parameters for arbitrary quantum systems}
\author{Michael te Vrugt}
\affiliation{Institut f\"ur Theoretische Physik, Center for Soft Nanoscience, Westf\"alische Wilhelms-Universit\"at M\"unster, D-48149 M\"unster, Germany}
	
\author{Raphael Wittkowski}
\email[Corresponding author: ]{raphael.wittkowski@uni-muenster.de}
\affiliation{Institut f\"ur Theoretische Physik, Center for Soft Nanoscience, Westf\"alische Wilhelms-Universit\"at M\"unster, D-48149 M\"unster, Germany}

\begin{abstract}		
The concept of quantum-mechanical nematic order, which is important in systems such as superconductors, is based on an  analogy to classical liquid crystals, where order parameters are obtained through orientational expansions. We generalize this method to quantum mechanics based on an expansion of Wigner functions. This provides a unified framework applicable to arbitrary quantum systems. The formalism is demonstrated for the cases of Fermi liquids and spin systems. Moreover, we derive new order parameters for molecular systems, which cannot be properly described with the usual nematic tensors.
\end{abstract}
\maketitle

The orientational order of classical liquid crystals is usually measured via Cartesian order-parameter fields given by the local polarization $\vec{P}(\vec{r})$ and nematic tensor $Q(\vec{r})$. These are typical variables of field theories for liquid crystals and active soft matter \cite{Pearce2019,MuellerYD2019,HartmannSPMSDDD2019,DoostmohammadiIYS2018,EllisPCGGF2018,YuanMSTS2018,StenhammarWMC2016,WittmannMM2016,PraetoriusVWL2018,OphausGT2018,WittkowskiLB2010,WittkowskiLB2011,WittkowskiLB2011b}. The standard definitions of $\vec{P}(\vec{r})$ and $Q(\vec{r})$ assume uniaxial (e.g., rodlike) particles and have to be modified for particles with low symmetry \cite{Rosso2007,LuckhurstS2015,MatsuyamaAWF2019}. This modification allows to study a much richer phenomenology of phase transitions \cite{GlotzerS2007,MatsuyamaAWF2019}. Classical order parameters can be defined systematically by an expansion of the orientational distribution function in symmetric traceless tensors \cite{EhrentrautM1998,Turzi2011,teVrugtW2019}, where $\vec{P}(\vec{r})$ and $Q(\vec{r})$ correspond to orders $l=1$ and $l=2$, respectively. Interesting phases can also be found using order parameters with $l \geq 3$ \cite{JakliLS2018,LubenskyR2002}.

In recent years, there has been an increasing interest in applying the concept of liquid-crystalline phases to quantum-mechanical systems \cite{PfauEtAl2019,HashimotoOYSSWCKMSOS2018,KangFC2018,LiuLHLWWSGZL2018,HardyHWWSMBAEH2019,HanaguriIKMWKSM2018,HuangY2018,FernandesCS2014,ChenEtAl2019,KonigP2018,ZiboldCFIDG2016,ZanLHWL2018,TsunetsuguA2006,BhattacharjeeSS2006,LiuK2019,DienerH2006,SongSZ2007,HamleyGHBC2012,OganesyanKF2001,FradkinKO2007,BaekEOKVJB2015,NakayamaMPSTUTT2014,SiYA2016,HarterZYMH2017,OrlovaLGCKHKR2017,HirobeSHSMS2019,BoudjadaWP2018,FradkinK1999,KivelsonFE1998,Read1989,WatanabeY2018,ColdeaBKHAWKCMRT2019,LeeSKHR2018,Shimojima2SNMKSMISI019,LicciardelloBLAKMSH2019,LiuLHLWWSGZL2018}. A source of nematic order in fermionic systems are deformations of the Fermi surface, which have gained significant attention in the past years \cite{ColdeaBKHAWKCMRT2019,LeeSKHR2018,Shimojima2SNMKSMISI019,LicciardelloBLAKMSH2019}. Nematic order is very important in superconductors \cite{PfauEtAl2019,HashimotoOYSSWCKMSOS2018,KangFC2018,LiuLHLWWSGZL2018,HardyHWWSMBAEH2019,HanaguriIKMWKSM2018,HuangY2018,FernandesCS2014,ChenEtAl2019}. It also arises in spin systems \cite{KonigP2018,ZanLHWL2018,TsunetsuguA2006,BhattacharjeeSS2006,LiuK2019,DienerH2006,SongSZ2007,HamleyGHBC2012,ZiboldCFIDG2016}, where it is relevant for ultracold gases \cite{ZiboldCFIDG2016,ZanLHWL2018,KonigP2018} and quantum computers \cite{KohamaTMKSH2019,KampermannV2002,SinhaMRK2001}.

Defining quantum-mechanical orientational order parameters systematically is a challenging problem, since quantum nematic order exists in various forms, such as structural, spin, and orbital. The relations between these forms are not fully understood \cite{FernandesCS2014,ChenEtAl2019,PfauEtAl2019} and it is difficult to connect the existing order parameters to their classical relates \cite{FradkinKO2007}. These problems could be solved through a systematic derivation of order parameters for quantum systems that naturally generalizes the classical case and unifies all known types of quantum order into one framework. This would allow to describe phases with $l \geq 3$, which occur, e.g., in Fermi fluids \cite{SunF2008,FradkinKLEM2010}, using Cartesian order parameters, and to improve the analysis of nematic order in systems such as superconductors where suitable order parameters are difficult to identify \cite{FernandesCS2014,PustogowLCSSMHRBB2019}. Other systems, where Cartesian quantum order parameters could then be derived, include molecules with low symmetry, where the usual nematic tensors are not sufficient \cite{Rosso2007,MatsuyamaAWF2019,LuckhurstS2015}. This is interesting, because molecular order plays a key role in chemical reactions \cite{Blum2012} and quantum dot technology \cite{MundoorSPASvdL2018,BasuI2009}. Like formalisms for different types of quantum order \cite{FurukawaMO2006}, this method would also allow to predict new phases for other quantum systems, which would be extremely useful for applications in solid-state theory, molecular physics, and beyond.

In this article, we derive a general formalism for the definition of quantum-mechanical orientational order parameters based on Wigner functions \cite{Wigner1932,Moyal1949,Case2008,TilmaESMN2016,McconnellZHV2015,BatemanNHU2014,WeinbubF2018}. These functions allow for a phase-space description of quantum systems by generalizing the classical probability distribution function and are widely used in fields such as solid-state theory \cite{HahnGKW2019,WiggerGARK2016}, quantum molecular dynamics \cite{SticklerSH2018,Filinov2008}, quantum information theory \cite{McconnellZHV2015,ArkhipovBS2018}, quantum optics \cite{SchulteHJMLA2015}, quantum electronics \cite{WoloszynS2017,WeinbubF2018}, plasma physics \cite{LarkinF2018}, and particle physics \cite{ProkhorovTZ2019,GaoLWW2018}. An expansion of the angular dependence of Wigner functions and corresponding kernel operators (see below) allows for a systematic definition of order parameters for the quantum case. Since Wigner functions can be constructed for arbitrary quantum systems \cite{TilmaESMN2016}, this approach has the major advantage that it is very general, being applicable to order arising from electronic structure, spin, or otherwise. Therefore, it solves the problem of a unification of the various types of existing quantum order parameters. Our approach has a clear connection to the classical case of which it is a natural extension. Moreover, performing an orientational expansion of Wigner functions for systems where nematic states are currently unknown would allow to describe and predict new phases. The standard order parameters for spin and Fermi nematics can be derived within our framework, which also allows to identify the approximations involved in their definitions. More general order parameters for molecules can be obtained in analogy to their classical relates.

In a classical liquid crystal, the orientation of a uniaxial particle is specified by an orientation vector $\uu(\theta,\phi)$, where $\theta$ and $\phi$ are spherical coordinate angles. (In two spatial dimensions, one angle $\phi$ is sufficient.) This vector specifies the direction of the particle's symmetry axis or, for an active particle, of its self-propulsion \cite{BickmannW2019,BickmannW2019b}. A many-particle system is then described by a distribution function $f(\uu)$ that in general also has other degrees of freedom, such as position or momentum. In three spatial dimensions, it can be expanded in spherical harmonics $Y_{lm}(\theta,\phi)$, giving the angular expansion
\begin{equation}
f(\theta,\phi)=\SUM{l=0}{\infty}\SUM{m=-l}{l} f_{lm} Y_{lm}(\theta,\phi)
\label{sphericalharmonics}%
\end{equation}
with coefficients $f_{lm}=\INTOII f(\theta,\phi) Y^\star_{lm}(\theta,\phi)$. Alternatively, one can perform a Cartesian expansion \cite{JoslinG1983,GrayG1984,teVrugtW2019}
\begin{equation}
f(\uu)=\SUM{l=0}{\infty} \SUM{i_{1},\dotsc,i_{l}=1}{d} \ff{d}{l}{i_{1}\dotsb i_{l}} u_{i_{1}}\!\dotsb u_{i_{l}},
\label{descartes}%
\end{equation}
where the expansion coefficients are given by 
\begin{equation}
\ff{d}{l}{i_{1}\dotsb i_{l}} = A^{(d\mathrm{D})}_{l} \INT{S_{d-1}}{}{\Omega} f(\uu) \TT{d}{l}{i_{1}\dotsb i_{l}}(\uu)
\label{tensor}%
\end{equation}
with the normalization $A^{(d\mathrm{D})}_{l}$ and tensor polynomials $\TT{d}{l}{i_{1}\dotsb i_{l}}(\uu)$ that depend on the spatial dimensionality $d$ (see Ref.\ \cite{teVrugtW2019} for definitions). We integrate over $\phi$ for $d=2$ and over $\phi$ and $\theta$ for $d=3$. The expansion coefficients \eqref{tensor} form symmetric traceless tensors being the orientational order parameters of order $l=0,1,\dotsc$. Angular and Cartesian expansions are orderwise equivalent.
	
The orientation of a particle with arbitrary shape is specified by a rotation matrix $R_{ij}$ that maps from a laboratory-fixed frame to a body-fixed frame. One can expand the distribution function $f(R)$ in various ways \cite{Rosso2007,LuckhurstS2015,Turzi2011}. An angular expansion can be performed in Wigner D-matrices $D^l_{mn}(R)$ \cite{GrayG1984} as a generalization of \cref{sphericalharmonics}. One possibility for a Cartesian expansion is, as shown by Turzi \cite{Turzi2011}, given by
\begin{equation}
f(R) = \sum_{l=0}^{\infty} \sum_{i_{1},\dotsc,i_{l}=1}^{3} \sum_{j_{1},\dotsc,j_{l}=1}^{3} \!\!\!\!\!\! c^{(3\mathrm{D})}_{i_1 j_1 \dotsb i_l j_l}R_{i_1 j_1} \!\dotsb R_{i_l j_l}.
\label{moregeneral}\raisetag{1em}%
\end{equation}
The expansion coefficients $c^{(3\mathrm{D})}_{i_1 j_1 \dotsb i_l j_l}$ (see Refs.\ \cite{teVrugtW2019,Turzi2011}) are symmetric and traceless in the $\{i_k\}$ and $\{j_k\}$ separately. Details on angular and Cartesian orientational expansions can be found in Ref.\ \cite{teVrugtW2019}.
	
Usually, the expansion \eqref{descartes} is performed up to second order. In two spatial dimensions, this gives \cite{WittkowskiSC2017,teVrugtW2019}
\begin{equation}
f(\uu) = f_0 + \sum_{i=1}^{2}P_i u_i + \sum_{i,j=1}^{2}Q_{ij}u_i u_j +\mathcal{O}(u^3)
\label{classical}%
\end{equation}
with the constant $f_{0}=\frac{1}{2\pi}\TINT{0}{2\pi}{\phi}f(\uu)$, polarization $P_i = \frac{1}{\pi}\TINT{0}{2\pi}{\phi}f(\uu) u_i$, and nematic tensor $Q_{ij} = \frac{2}{\pi}\TINT{0}{2\pi}{\phi}f(\uu)(u_i u_j - \frac{1}{2}\delta_{ij})$. 
Using \cref{descartes,moregeneral}, generalizations to three spatial dimensions, higher orders, and asymmetric particles are straightforward. 
	
It is not immediately clear how this formalism can be applied to quantum mechanics. The first difference is that, while classical order parameters for passive systems measure the geometric orientation of molecules, quantum nematic order arises in systems of point-like particles \cite{FradkinKLEM2010}. (One can find here an interesting analogy between active and quantum matter, since active spheres can also show nematic order despite being geometrically symmetric \cite{WittkowskiSC2017,teVrugtW2019}.) The second difference is that quantum systems are usually described using a statistical operator $\hat{\rho}$, which is not defined on a classical phase space, so that there is no $\uu$-dependence that could be expanded.
	
A solution to this problem is provided by a phase-space representation of a quantum system. Such representations can be developed in various forms. We here use Wigner functions due to their close relation to the classical probability distribution function \cite{TilmaESMN2016,HilleryOSW1984}. It is expanded in the same way. The expansion coefficients provide order parameters for quantum systems. By transforming back to the operator formalism, one then obtains order parameters in the form of Hermitian Hilbert space operators, constituting suitable observables for quantum systems.

Formally, a function $W(\vec{\Gamma})$ depending on the phase-space coordinates $\vec{\Gamma}$ is a Wigner function for a Hilbert space operator $\hat{\rho}$ if there exists a Hermitian kernel operator $\hat{\Delta}(\vec{\Gamma})$, the Stratonovich-Weyl kernel, such that \cite{Stratonovich1957,BrifM1999,TilmaESMN2016}
\begin{gather}
W(\vec{\Gamma}) = \Tr(\hat{\rho}\hat{\Delta}(\vec{\Gamma})),\\
\hat{\rho} = \INT{}{}{\Gamma} W(\vec{\Gamma})\hat{\Delta}(\vec{\Gamma})
\label{rhohat}%
\end{gather}
with the quantum-mechanical trace $\Tr$. This, along with a few additional requirements (see Refs.\ \cite{BrifM1999,TilmaESMN2016} for details), is the Stratonovich-Weyl correspondence. More generally, \cref{rhohat} can be thought of as a quantization rule \cite{AbromHS2000}: The Hilbert space operator $\hat{A}$ corresponding to a phase-space function $A(\vec{\Gamma})$ is \cite{BrifM1999}
\begin{equation}
\hat{A} = \INT{}{}{\Gamma} A(\vec{\Gamma})\hat{\Delta}(\vec{\Gamma}).
\label{quantization}%
\end{equation}
	
Now suppose that our system has orientational degrees of freedom, i.e., $W(\vec{\Gamma}) \equiv W(\uu,\vmm)$ with $\vmm$ containing possible nonorientational degrees of freedom (that can be integrated out if necessary).  We can expand a Wigner function using \cref{descartes} as
\begin{equation}
W(\vec{\Gamma}) = \SUM{l=0}{\infty} \SUM{i_{1},\dotsc,i_{l}=1}{d} W_{i_{1}\dotsb i_{l}}(\vmm) u_{i_{1}} \!\dotsb u_{i_{l}}.
\label{wignerexpansion}
\end{equation}
The prefactors $W_{i_{1}\dotsb i_{l}}(\vmm)$ provide order parameters for calculations in the Wigner function formalism. They can be easily calculated using \cref{tensor}. Since Wigner functions can be measured \cite{LutterbachD1997,KurtsieferPM1997}, the order parameters defined in this way are experimentally accessible quantities.

Alternatively, we can define order parameters in terms of Hermitian Hilbert space operators, which can be used as observables in \ZT{ordinary} quantum mechanics. This can be achieved by expanding a kernel operator as
\begin{equation}
\INT{}{}{\mm}\hat{\Delta}(\uu,\vmm) = \SUM{l=0}{\infty} \SUM{i_{1},\dotsc,i_{l}=1}{d} u_{i_{1}} \!\dotsb u_{i_{l}} \hat{T}_{i_{1}\dotsb i_{l}}
\label{deltaexpansion}%
\end{equation}
with the quantum-mechanical order parameters
\begin{equation}
\hat{T}_{i_{1}\dotsb i_{l}} = A^{(d\mathrm{D})}_{l} \INT{S_{d-1}}{}{\Omega}\INT{}{}{\mm} \TT{d}{l}{i_{1}\dotsb i_{l}}(\uu) \hat{\Delta}(\uu,\vmm),
\label{general}%
\end{equation}
where the normalization $A^{(d\mathrm{D})}_{l}$ is chosen in such a way that \cref{descartes,tensor} hold for the invariant integration measure $\dif\Gamma=\dif\Omega\dif\mm$. Here, we need to integrate out also the nonorientational degrees of freedom such that the operators $\hat{T}_{i_{1}\dotsb i_{l}}$ have no phase-space dependence, making them proper quantum observables. Since $\hat{\Delta}(\vec{\Gamma})$ is Hermitian \cite{TilmaESMN2016}, the $\hat{T}_{i_{1}\dotsb i_{l}}$ are as well.

Comparing \cref{quantization,general} shows that \cref{general} is a quantization prescription.  Hence, the quantum-mechanical Cartesian order parameters $\hat{T}_{i_{1}\dotsb i_{l}}$ are obtained by a quantization of the normalized tensor polynomials $A^{(d\mathrm{D})}_{l}\TT{d}{l}{i_{1}\dotsb i_{l}}(\uu)$.  When we apply the general definition \eqref{general} to specific systems, the definition of the order parameters depends on the kernel operator $\hat{\Delta}(\uu,\vmm)$. Such a kernel can be constructed for any quantum system by using its underlying symmetries \cite{TilmaESMN2016,BrifM1999}. 

Finally, we need to take into account that we are typically dealing with many-particle systems. By integrating over the coordinates of all particles except for one, we can obtain a reduced one-particle Wigner function that allows for an orientational expansion \cite{CancellieriBJ2007}. In \cref{deltaexpansion}, we replace for $N$ particles 
\begin{equation}
\hat{\Delta}(\vec{\Gamma}) \to \frac{1}{N}\SUM{i=1}{N} \bigg( \prod_{\begin{subarray}{c}j=1\\j\neq i\end{subarray}}^{N}\INT{}{}{\Gamma_j} \bigg) \hat{\Delta}(\{\vec{\Gamma}_{k}\}),
\label{deltamanyparticle}%
\end{equation}
which ensures that all particles are treated equally. Since the kernel for an ensemble of $N$ particles is constructed from a tensor product of the individual kernels \cite{RundleTSDBE2019,KoczorZG2019}, the resulting operators $\hat{T}_{i_{1}\dotsb i_{l}}$ act on the $N$-particle Hilbert space as they should.

As a first specific system, we consider Fermi liquids. Here, nematic states break a rotational symmetry of the underlying crystal. In the simplest case of a system with full rotational symmetry, fermions fill up a ball of radius $p_F$ (Fermi momentum) in reciprocal space, whose surface is referred to as the Fermi surface. Nematic order then corresponds to a state in which the Fermi surface becomes elliptical due to a thermodynamic instability \cite{FradkinKLEM2010,OganesyanKF2001}.

We describe fermions using Wigner functions depending on position $\vec{x}$ and momentum $\vec{p}$. When setting $\hbar =1$ and $\dif\Gamma = (2\pi)^{-2}\dif^2 x\dif^2 p$, the kernel for a system with phase-space coordinates $\vec{\Gamma}=(\vec{x},\vec{p})^{\mathrm{T}}$ in two spatial dimensions reads \cite{WaalkensSW2007,GneitingFH2013,Weyl1927}
\begin{equation}
\hat{\Delta}(\vec{x},\vec{p})= \frac{1}{(2\pi)^{2}}\INT{}{}{^2\xi}\INT{}{}{^2\zeta}e^{-\ii(\vec{\xi}\cdot(\vec{x}-\hat{\vec{x}})+\vec{\zeta}\cdot(\vec{p}-\hat{\vec{p}}))}
\label{fermionkernel}%
\end{equation}
with position operator $\hat{\vec{x}}$ and momentum operator $\hat{\vec{p}}$. We are interested in deviations from spherical symmetry in reciprocal space. Therefore, we write $\vec{p} = p \uu$ with $p = \norm{\vec{p}}$ and then perform an orientational expansion. Applying the general definition \eqref{general} together with the nematic tensor $Q_{ij}$ from \cref{classical} (multiplied by $(2\pi)^2$ for normalization) and \cref{fermionkernel} gives 
\begin{equation}
\begin{split}
\hat{Q}_{ij} &= \INT{}{}{^2x}\INT{}{}{^2p}(2u_i u_j - \delta_{ij})\hat{\Delta}(\vec{x},p,\uu)\\
&= \frac{1}{(2\pi)^{4}}\INT{}{}{^2x}\INT{}{}{^2p}\INT{}{}{^2\xi}\INT{}{}{^2\zeta}\\
&\quad\;\:\!\frac{(2\pi)^{2}}{p^2}(2p_i p_j - p^2\delta_{ij}) e^{-\ii(\vec{\xi}\cdot(\vec{x}-\hat{\vec{x}})+\vec{\zeta}\cdot(\vec{p}-\hat{\vec{p}}))}.
\end{split}
\end{equation}
For $N$ particles, using \cref{deltamanyparticle} and $\TINT{}{}{\Gamma}\hat{\Delta}(\vec{\Gamma}) = \mathds{1}$ \cite{TilmaESMN2016} gives $\hat{Q}_{ij}$ as a sum over the single-particle order parameters, normalized with $1/N$. If the deviations from spherical symmetry are small, only states close to the Fermi surface contribute to the expectation value $\braket{\hat{Q}_{ij}}$ and we can approximate $1/p^2 \approx 1/p_F^2$. This allows us, if we switch to matrix notation, to write
\begin{equation}
\begin{split}
\hat{Q}&=\frac{1}{(2\pi)^{4}}\INT{}{}{^2x}\INT{}{}{^2p}\INT{}{}{^2\xi}\INT{}{}{^2\zeta} \\
&\quad\;\:\! \frac{(2\pi)^{2}}{p_F^2}
\begin{pmatrix}
p_x^2 - p_y^2& 2p_x p_y\\
2p_x p_y & p_y^2 - p_x^2\\
\end{pmatrix} e^{-\ii(\vec{\xi}\cdot(\vec{x}-\hat{\vec{x}})+\vec{\zeta}\cdot(\vec{p}-\hat{\vec{p}}))}.
\end{split}
\label{comp}\raisetag{3.5em}%
\end{equation}	
Comparing \cref{comp,quantization} shows that \cref{comp} is just the Weyl quantization of the matrix. Here, it can be evaluated with the replacement $p_i \to \hat{p}_i = -\ii\partial_i$ \cite{Hall2013,WaalkensSW2007,GneitingFH2013}, yielding
\begin{equation}
\begin{split}
\hat{Q}=&-\frac{(2\pi)^{2}}{p_F^2}
\begin{pmatrix}
\partial_x^2 - \partial_y^2& 2\partial_x \partial_y\\
2\partial_x \partial_y & \partial_y^2 - \partial_x^2\\
\end{pmatrix}.\\
\end{split}
\end{equation}
This is, up to a normalization resulting from the integral measure, the definition of the nematic order parameter for Fermi liquids used in the literature  \cite{OganesyanKF2001,Fradkin2012}. The general definition \eqref{general} thus reproduces the standard definitions if it is applied to standard systems.
	
Quantum-mechanical nematic order also arises in systems of spins. Spins can be described using Wigner functions that depend on the angles $\theta$ and $\phi$, i.e., we have $\vec{\Gamma} = \uu$. The kernel operator for spins is given by \cite{Klimov2002,KalmykovCT2008}
\begin{equation}
\hat{\Delta}(\theta,\phi) = \sqrt{\frac{4\pi}{2s+1}}\sum_{l=0}^{2s}\sum_{m=-l}^{l}Y^\star_{lm}(\theta,\phi)\hat{T}^{(s)}_{lm}
\label{spinkernel}%
\end{equation}
with the irreducible tensor operators \cite{KalmykovCT2008,delaHozBKLS2014,RundleTSDBE2019}
\begin{equation}
\hat{T}^{(s)}_{lm}=\sqrt{\frac{2l+1}{2s+1}}\sum_{n,n'=-s}^{s}C^{s,n'}_{s,n,l,m}\ket{s,n'}\bra{s,n}
\label{tensoroperators}%
\end{equation}
and the Clebsch-Gordan coefficients $C^{s,n'}_{s,n,l,m}$. 

Since the kernel \eqref{spinkernel} can be easily brought into the form \eqref{sphericalharmonics} by conjugation, the simplest way to obtain a Cartesian expansion is to read off the angular expansion coefficients and convert them to Cartesian ones (see Ref.\ \cite{teVrugtW2019} for conversion tables). The tensor operators \eqref{tensoroperators} can be expressed in terms of the spin operator $\hat{\vec{S}}$ \cite{VarshalovichMK1988,BuckmasterCS1972,BowdenH1986,delaHozBKLS2014,ChenCZ2016,Blum2012}. For a general spin $s$, we find 
\begin{align}
\hat{P}_i = \frac{3}{\sqrt{s(s+1)(2s+1)^2}}\hat{S}_i
\label{polspin}%
\end{align}
for the polarization and
\begin{equation}
\begin{split}
\hat{Q}_{ij} &= \frac{15}{2}\frac{1}{\sqrt{s(s+1)(2s-1)(2s+1)^2(2s+3)}}\\
&\quad\;\:\! \Big(\hat{S}_i\hat{S}_j + \hat{S}_j\hat{S}_i-\frac{2}{3}s(s+1)\delta_{ij}\hat{I}\Big)
\end{split}
\label{nematicspin}%
\end{equation}
with the identity operator $\hat{I}$ for the nematic tensor. In case of $N$ particles, applying \cref{deltamanyparticle} with the single-particle kernel \eqref{spinkernel} gives the many-particle order parameter as a normalized sum over the single-particle order parameters. These definitions agree, for $s=1$, up to a normalization with the standard definitions of the polarization \cite{Blum2012} and nematic tensor \cite{VarshalovichMK1988,BhattacharjeeSS2006,ChenCZ2016}. They are usually obtained by a state multipole expansion of the density operator \cite{Blum2012}, a procedure that is less general than the formalism discussed here, since it is restricted to angular momenta. Thus, we can again recover the definitions used in the literature, but also generalize them towards larger spins. Equation \eqref{general}, which is our general result, therefore contains both important types of order -- spin and Fermi nematics -- as special cases.

The definition \eqref{nematicspin} is very similar to its classical counterpart from \cref{classical}, except for the fact that it is symmetrized because the components of the spin operator do not commute.  What is also interesting here is that, while the classical order-parameter expansions \eqref{sphericalharmonics} and \eqref{descartes} go from $l=0$ to $l=\infty$, the expansion for spin systems stops at $l=2s$. Consequently, since nematicity is measured by the terms of order $l=2$, there can be no nematic order in spin-1/2 systems. The physical reason is that spin-1/2 particles are always locally magnetic (except if they couple to a spin-1 system, which is possible, e.g., for magnons) \cite{Mila2017,ZhitomirskyT2010,Blum2012}. Indeed, spin nematic states have only been found in spin-1-systems \cite{OrlovaLGCKHKR2017,BhattacharjeeSS2006,Mila2017,ZhitomirskyT2010,TsunetsuguM2007,AndreevG1984}. Within our framework, this naturally follows from the general theory of order parameters, which also allows to define higher-order phases for spin-3/2 particles that are known to have richer phase diagrams \cite{Wu2005}. Note that the relations used here assume a sharp angular momentum \cite{Blum2012}. In other cases, the form of the order parameters might change, which can also be calculated within our formalism: Wigner functions for systems where various SU(2)-invariant subspaces are relevant depend on three angles \cite{KlimovR2008}, so that we require a more general expansion.

The situation is similar for molecules: For a specification of the orientation of classical molecules with low symmetry, one requires a rotation $R$ rather than an orientation vector $\uu$. Hence, the molecules are described using distribution functions $f(R)$ rather than $f(\uu)$. In consequence, order parameters for these systems require more general expansions such as \eqref{moregeneral} to give a complete description of the orientational state \cite{Rosso2007,MatsuyamaAWF2019,LuckhurstS2015,LuckhurstS2015}. A general quantum-mechanical molecule can be modelled as a quantum rigid rotor \cite{Blum2012,MoralesDO1999}, which is described using Wigner functions $W(R)$ (see, e.g., Refs.\ \cite{FischerGH2013,RomeroKW2015} for explicit forms). Nematic order in quantum molecular systems can therefore not be described with the usual methods, such as the construction of symmetric traceless tensors used for Fermi liquids \cite{FradkinKLEM2010} and spins \cite{ChenCZ2016}. The typical approach for the definition of order parameters for such molecular systems is through angular averages \cite{Blum2012}, although Cartesian order parameters are more closely linked to experiments \cite{Turzi2011,teVrugtW2019}. A more appropriate description is possible using our approach. Applying the expansion \eqref{moregeneral} to the kernel $\hat{\Delta}(R)$ for a molecular system gives the operators for polarization \cite{teVrugtW2019}
\begin{equation}
\hat{P}_{ij} = \frac{3}{8\pi^2}\INT{\mathrm{SO(3)}}{}{\Omega}R_{ij}\hat{\Delta}(R)
\end{equation}
and nematic order
\begin{equation}
\begin{split}
\hat{Q}_{ijkl} &= \frac{5}{16\pi^2}\INT{\mathrm{SO(3)}}{}{\Omega} \Big(R_{ij}R_{kl} + R_{il}R_{kj}
 - \frac{2}{3}\delta_{ik}\delta_{jl}\Big)\hat{\Delta}(R),
\end{split}\raisetag{0.7em}
\end{equation}
where the integral goes over all possible rotations.

In summary, we have derived a systematic definition of orientational order parameters for arbitrary quantum systems based on Cartesian expansions of Wigner functions and Stratonovich-Weyl kernels. Our definition can recover the standard definitions for both Fermi liquids and spins, even though there are very different physical mechanisms involved. Moreover, our framework allows to define new order parameters, as demonstrated for the case of molecular systems. This unifying framework allows for a better understanding of quantum order parameters and gives the opportunity to discover nematic phases in systems where they could not be defined previously. The order parameters for low-symmetry molecules can help to extend the study of phase transitions in systems of particles with low or no symmetry towards quantum systems. Finally, as in classical soft matter physics \cite{LubenskyR2002}, order parameters with $l \geq 3$ can allow to find new types of phases.
	
We thank Tilmann Kuhn and Jonas L\"ubken for helpful discussions. 
R.W.\ is funded by the Deutsche Forschungsgemeinschaft (DFG, German Research Foundation) -- WI 4170/3-1.

\nocite{apsrev41Control}
\bibliographystyle{apsrev4-1}
\bibliography{control,refs}

\end{document}